\documentclass{article}
\usepackage{nips10submit_e,times}
\usepackage{graphicx}
\usepackage{amssymb}
\usepackage{amsmath}

\newcommand*{\set}[1]{\ensuremath{\mathcal{#1}}}
\renewcommand*{\vec}[1]{\boldsymbol{#1}}
\newcommand{\Reals}{{\mathbb{R}}}

\nipsfinalcopy 

\title{Currency Forecasting using Multiple Kernel Learning with Financially Motivated Features}

\author{
Tristan Fletcher, Zakria Hussain and John Shawe-Taylor\\
Centre for Computational Statistics and Machine Learning\\
Department of Computer Science\\
University College London, UK\\
\texttt{\{t.fletcher,z.hussain,jst\}@cs.ucl.ac.uk} \\
}
\begin{document}

\maketitle

\begin{abstract}
Multiple Kernel Learning (MKL) is used to replicate the signal combination process that trading rules embody when they aggregate multiple sources of financial information when predicting an asset's price movements. A set of financially motivated kernels is constructed for the EURUSD currency pair and is used to predict the direction of price movement for the currency over multiple time horizons. MKL is shown to outperform each of the kernels individually in terms of predictive accuracy. Furthermore, the kernel weightings selected by MKL highlights which of the financial features represented by the kernels are the most informative for predictive tasks. \footnote{This work is closely related to a presentation titled \textit{Multiple Kernel Learning on the Limit Order Book} given at the Workshop on Applications of Pattern Analysis 2010.} 
\end{abstract}
\section{Introduction}

\noindent A trader wishing to speculate on a currency's movement is most interested in what direction he believes the price of that currency $P_{t}$ will move over a time horizon $\Delta t$ so that he can take a position based on this prediction. Any move that is predicted has to be significant enough to cross the difference between the buying price (bid) and selling price (ask) in the appropriate direction if the trader is to profit from it. If we view this as a three class classification task, then we can simplify this aim into an attempt to predict whether the trader should buy the currency pair because he believes $P^{Bid}_{t+\Delta t}>P^{Ask}_{t}$, sell it because $P^{Ask}_{t+\Delta t}<P^{Bid}_{t}$ or do nothing because $P^{Bid}_{t+\Delta t}<P^{Ask}_{t}$ and $P^{Ask}_{t+\Delta t}>P^{Bid}_{t}$. 

When making trading decisions such as whether to buy or sell a currency, traders typically combine the information from many models to create an overall trading rule (see for example \cite{Kaufman2005}). The aim of this work is to represent this model combination process through Multiple Kernel Learning, where individual kernels based on common trading signals are created to represent the constituent sources of information. 

There has been much work in using kernel based methods such as the SVM to predict the movement of financial time series, e.g. \cite{Tay2001}, \cite{Tay2002}, \cite{Cao2003}, \cite{Kim2003}, \cite{Perez2003}, \cite{Huang2005}, \cite{Gestel2001} ,\cite{Hazarika2002}, \cite{Tino2005}, \cite{Huang2006}, \cite{Huang2008}, \cite{Chalup2008}, \cite{Fletcher2009} and \cite{Ullrich2009}. However the majority of the previous work in this area deals with the problem of kernel selection in a purely empirical manner with little to no theoretical justification, and makes no attempts to either use financially plausible kernels or indeed to combine kernels in any manner.

\section{Financially Motivated Features}

\subsection{Price-based Features}

\noindent The following four features are based on common price-based trading rules (which are described briefly in the Appendix):

$\set{F}_{1}=\left\{EMA^{L_{1}}_{t}, \ldots, EMA^{L_{N}}_{t} \right\}$

$\set{F}_{2}=\left\{MA^{L_{1}}_{t}, \ldots, MA^{L_{N}}_{t}, \sigma^{L_{1}}_{t}, \ldots, \sigma^{L_{N}}_{t} \right\}$ 

$\set{F}_{3}=\left\{P_{t},\max_{t}^{L_{1}}, \ldots, \max_{t}^{L_{N}}, \min_{t}^{L_{1}}, \ldots, \min_{t}^{L_{N}} \right\}$

$\set{F}_{4}=\left\{\Uparrow_{t}^{L_{1}}, \ldots, \Uparrow_{t}^{L_{N}}, \Downarrow_{t}^{L_{1}}, \ldots, \Downarrow_{t}^{L_{N}} \right\}$

\noindent where $EMA^{L_{i}}_{t}$ denotes an exponential moving average of the price $P$ at time $t$ with a half life $L_{i}$, $\sigma^{L_{i}}_{t}$ denotes the standard deviation of $P$ over a period $L_{i}$, $MA^{L_{i}}_{t}$ its simple moving average over the period $L_{i}$, $\max_{t}^{L_{i}}$ and $\min_{t}^{L_{i}}$ the maximum and minimum prices over the period and $\Uparrow_{t}^{L_{i}}$ and $\Downarrow_{t}^{L_{i}}$ the number of price increases and decreases over it.

\subsection{Volume-based Features}

\noindent The majority of currency trading takes place on Electronic Communication Networks (ECNs). Continuous trading takes place on these exchanges via the arrival of limit orders specifying whether the party wishes to buy or sell, the amount (volume) desired, and the price the transaction will occur at. While traders had previously been able to view the prices of the highest buy (best bid) and lowest sell orders (best ask), a relatively recent development in certain exchanges is the real-time revelation of the total volume of trades sitting on the ECN's order book at both these price levels and also at price levels above the best ask and below the best bid. This exposure of order books' previously hidden depths allows traders to capitalize on the greater dimensionality of data available to them when making trading decisions and suggests the use of kernel methods on this higher dimensional data.

\noindent Representing the volume at time $t$ at each of the price levels of the order book on both sides as a vector $\vec{V}_{t}$, where $\vec{V}_{t}\in\Reals^{6}$ for the case of three price levels on each side, a further set of four features can be constructed: 

$\set{F}_{5\ldots8}=\left\{\vec{V}_{t}, \frac{\vec{V}_{t}}{\left\|\vec{V}_{t}\right\|_{1}}, \vec{V}_{t}-\vec{V}_{t-1}, \frac{\vec{V}_{t}-\vec{V}_{t-1}}{\left\|\vec{V}_{t}-\vec{V}_{t-1}\right\|_{1}} \right\}$

\section{Experimental Design}

\noindent Radial Basis Function (RBF) and polynomial kernels have often been used in financial market prediction problems, e.g. \cite{Huang2005} and \cite{Ullrich2009}. Furthermore, Artificial Neural Networks (ANN) are often used in financial forecasting tasks (e.g. \cite{KuanLiu1995}, \cite{Walczak2001} and \cite{Shadbolt2002}) and for this reason a kernel based on Williams (1998) \cite{Williams1998} infinite neural network with a sigmoidal transfer function is also employed (see $\set{K}_{11:15}$ below). A feature mapping set consisting of 5 of each of these kernel types with different values of the relevant hyperparameter ($\sigma$, $d$ or $\Sigma$) along with the linear kernel is used:
\begin{align*}
\set{K}_{1:5} 		&=\left\{ \exp\left(-\left\|\vec{x}-\vec{x}'\right\|^{2} / \sigma_{1}^{2}\right),  		\ldots, \exp\left(-\left\|\vec{x}-\vec{x}'\right\|^{2} / \sigma_{5}^{2}\right) 				\right\}\\
\set{K}_{6:10}		&=\left\{ \left(\left\langle \vec{x},\vec{x}' \right\rangle + 1\right)^{d_{1}},				\ldots, \left(\left\langle \vec{x},\vec{x}' \right\rangle + 1\right)^{d_{5}}					\right\}\\
\set{K}_{11:15}	  &=\left\{ \frac{2}{\pi} \sin^{-1} \left( \frac{2\vec{x}^{T}\Sigma_{1}\vec{x}'}{\sqrt{(1+2\vec{x}^{T}\Sigma_{1}\vec{x})(1+2\vec{x}'^{T}\Sigma_{1}\vec{x}')}}\right),\ldots, \frac{2}{\pi} \sin^{-1}\left( \frac{2\vec{x}^{T}\Sigma_{5}\vec{x}'}{\sqrt{(1+2\vec{x}^{T}\Sigma_{5}\vec{x})(1+2\vec{x}'^{T}\Sigma_{5}\vec{x}')}} \right) 																									 \right\}\\
\set{K}_{16}			&=\left\{ \left\langle \vec{x},\vec{x}' \right\rangle																																																												\right\}
\end{align*}																																																											
																																																											
This means that altogether there are $\left|\set{F}\right| \times \left|\set{K} \right| = 8 \times 16 = 128$ feature / kernel combinations. We will adopt notation so that for example the combination $\set{F}_{1}\set{K}_{1}$ is the moving average crossover feature with a RBF using the scale parameter $\sigma_{1}^{2}$.

Three SVM are trained on the data with the following labeling criteria for each SVM:
\begin{align*}
&\mbox{\indent SVM 1:\indent}P^{Bid}_{t+\Delta t}>P^{Ask}_{t}&\Rightarrow y_{t}^{1}=+1 &\mbox{, otherwise } y_{t}^{1}=-1 \\
&\mbox{\indent SVM 2:\indent}P^{Ask}_{t+\Delta t}<P^{Bid}_{t}&\Rightarrow y_{t}^{2}=+1 &\mbox{, otherwise } y_{t}^{2}=-1 \\
&\mbox{\indent SVM 3:\indent}P^{Bid}_{t+\Delta t}<P^{Ask}_{t},P^{Ask}_{t+\Delta t}>P^{Bid}_{t}&\Rightarrow y_{t}^{3}=+1 &\mbox{, otherwise } y_{t}^{3}=-1 
\end{align*}
In this manner, a three dimensional output vector $\vec{y}_{t}$ is constructed from $y_{t}^{1}$, $y_{t}^{2}$ and $y_{t}^{3}$ for each instance such that $\vec{y}_{t}=[\pm1,\pm1,\pm1]$. Predictions are only kept for instances where exactly one of the signs in $\vec{y}_{t}$ is positive, i.e. when all three of the classifiers are agreeing on a direction of movement. For this subset of the predictions, a prediction is deemed correct if it correctly predicts the direction of spread-crossing movement (i.e. upwards, downwards or no movement) and incorrect if not.

The MKL method of SimpleMKL \cite{Rakotomamonjy2008} is investigated along with standard SVM based on each of the 128 kernels / feature combinations individually. Predictions for time horizons ($\Delta t$) of 5, 10, 20, 50, 100 and 200 seconds into the future are created. Training and prediction is carried out by training the three SVM on 100 instances of in sample data, making predictions regarding the following 100 instances and then rolling forward 100 instances so that the out of sample data points in the previous window become the current window's in sample set. The data consists of $6 \times 10^{4}$ instances of order book updates for the EURUSD currency pair from the EBS exchange starting on 2/11/2009. \footnote{EURUSD was selected as the currency pair to investigate because it is the world's most actively traded currency pair, comprising 27\% of global turnover \cite{BIS2007}. Consequently, the EBS exchange was selected for this analysis because it is the primary ECN for EURUSD.}

\section{Results and Conclusions}

When comparing the predictive accuracy of the kernel methods when used individually to their combination in MKL one needs to consider both how often each method was able to make a prediction as described above and how correct the predictions were overall for the whole dataset. In the tables and figures that follow, for the sake of clarity only three of the 128 individual kernels are used when comparing SimpleMKL to the individual kernels. 10-fold cross-validation was used to select the three kernels with the highest predictive accuracy for the dataset, namely $\set{F}_{8}\set{K}_{16}$, $\set{F}_{1}\set{K}_{1}$ and $\set{F}_{1}\set{K}_{3}$.

\begin{figure}
\begin{center}
{\includegraphics[width=1.0\linewidth]{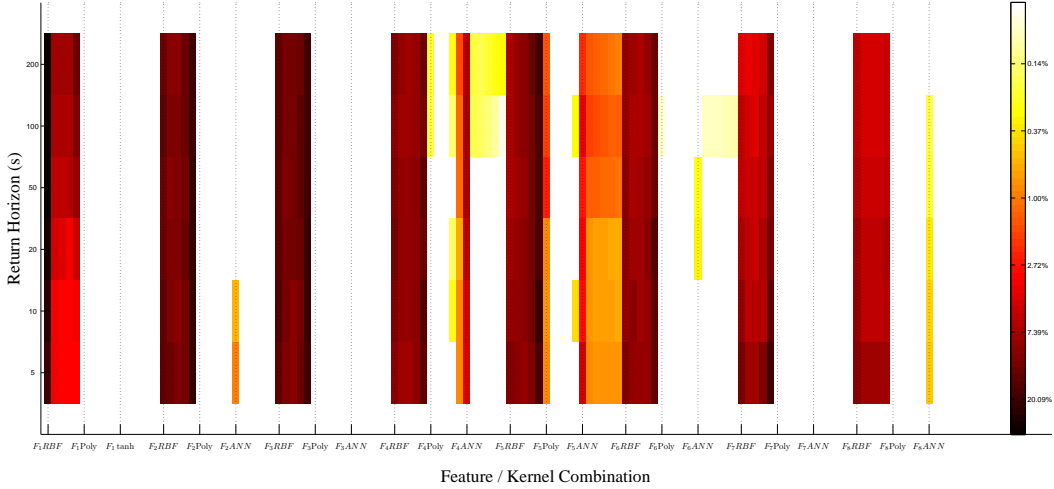}}
\label{KernelWeightsNIPS}
\end{center}
\caption{MKL Kernel weightings}
\end{figure}

\begin{table}[t]
\caption{Percentage of time predictions possible}
\label{PerPoss}
\begin{center}
\begin{tabular}{|l|c|c|c|c|}
\hline
\textbf{$\Delta t$}&\textbf{SimpleMKL}&$\set{F}_{8}\set{K}_{16}$&$\set{F}_{1}\set{K}_{1}$&$\set{F}_{1}\set{K}_{3}$\\\hline
\textbf{  5}&26.1&24.7&26.1&24.7\\\hline
\textbf{ 10}&41.1&40.4&39.8&37.7\\\hline
\textbf{ 20}&50.2&49.1&48.1&45.0\\\hline
\textbf{ 50}&46.3&44.1&44.8&45.5\\\hline
\textbf{100}&32.8&33.5&34.6&35.3\\\hline
\textbf{200}&27.0&24.9&26.6&27.4\\\hline
\end{tabular}
\end{center}
\end{table}

\begin{table}[t]
\caption{Percentage accuracy of predictions}
\label{PerAcc}
\begin{center}
\begin{tabular}{|l|c|c|c|c|}
\hline
\textbf{$\Delta t$}&\textbf{SimpleMKL}&$\set{F}_{8}\set{K}_{16}$&$\set{F}_{1}\set{K}_{1}$&$\set{F}_{1}\set{K}_{3}$\\\hline
\textbf{  5}&94.7&94.7&93.0&92.8\\\hline
\textbf{ 10}&89.9&89.6&88.4&84.6\\\hline
\textbf{ 20}&81.7&81.3&79.5&72.3\\\hline
\textbf{ 50}&67.1&65.4&65.5&61.1\\\hline
\textbf{100}&61.1&51.1&60.7&59.9\\\hline
\textbf{200}&58.9&45.0&58.8&61.3\\\hline
\end{tabular}
\end{center}
\end{table}

Table \ref{PerPoss}, which shows how often each of the methods were able to make a prediction for each of the time horizons, indicates that SimpleMKL was very similar in the frequency with which it was able to make predictions as the three individual kernel / feature combinations highlighted. Table \ref{PerAcc} shows each of the method's predictive accuracy over the entire dataset when a prediction was actually possible. The results indicate that SimpleMKL has higher predictive accuracy than the most effective individual kernels for all time horizons under $200$ seconds and is only marginally less effective than $\set{F}_{1}\set{K}_{3}$ for the $200$ second forecast horizon.

P-values for the null hypothesis that the results reported could have occurred by chance were calculated (the methodology for doing this is explained in the Appendix). It was found that for both SimpleMKL and the individual kernels highlighted for all forecast horizons, the null hypothesis could be rejected for a significance level of $<10^{-5}$.

As reflected in Figure 1, the kernel / feature combinations $\set{F}_{1}\set{K}_{1}$, $\set{F}_{2}\set{K}_{5}$ and $\set{F}_{3}\set{K}_{5}$ are consistently awarded the highest weightings by SimpleMKL and hence are the most relevant for making predictions over the data set. These kernels are the RBF mapping with the smallest scale parameter on the exponential moving average crossover feature, the RBF mapping with the largest scale parameter on the price standard deviation / moving average feature and the RBF mapping with the largest scale parameter again on the minimums / maximums feature. 

The vertical banding of colour (or intensity) highlights the consistency of each of the kernel / feature combination's weightings across the different time horizons: in almost all cases the weighting for a particular combination is not significantly different between when being used to make a prediction for a short time horizon and a longer term one. One can also see from Figure 1 that although all $8$ of the features have weightings assigned to them, in most cases this is only in conjunction with the RBF kernels - the polynomial (\textit{Poly}) and infinite neural network (\textit{ANN}) based mappings being assigned weightings by MKL for only the fourth and fifth features.

The most successful individual kernels as selected by cross-validation are awarded very low weights by SimpleMKL. This reflects a common feature of trading rules where individual signals can drastically change their significance in terms of performance when used in combination. Furthermore, the outperformance of SimpleMKL to the individual kernels highlighted indicates that MKL is an effective method for combining a set of price and volume based features in order to correctly forecast the direction of price movements in a manner similar to a trading rule.

\subsubsection*{Acknowledgments}
The authors would like to thank ICAP for making its EBS foreign exchange data available for this research.

\small
\bibliography{Fletcher}
\normalsize
\subsection*{Appendix}

\subsubsection*{Price-based Features}

\begin{itemize}
\item $\set{F}_{1}$: A common trading rule is the moving average crossover technique (see for example \cite{Appel2005}) which suggests that the price $P_{t}$ will move up when its short term moving average $EMA^{short}_{t}$ crosses above a longer term one $EMA^{long}_{t}$ and visa versa. 

\item $\set{F}_{2}$: Breakout trading rules (see for example \cite{Faith2007}) look to see if the price has broken above or below a certain threshold and assume that once the price has broken through this threshold the direction of the price movement will persist. One way of defining this threshold is through the use of Bollinger Bands \cite{Bollinger2001} where the upper/lower thresholds are set by adding/subtracting a certain number of standard deviations of the price movement $\sigma^{L}_{t}$ to the average price $MA_{L}^{t}$ for a period $L$.

\item $\set{F}_{3}$: Another breakout trading rule called the Donchian Trend system \cite{Faith2007} determines whether the price has risen above its maximum $\max_{t}^{L}$ or below its minimum $\min_{t}^{L}$ over a period $L$ and once again assumes that once the price has broken through this threshold the direction of the price movement will persist.

\item $\set{F}_{4}$: The Relative Strength Index trading rule \cite{Wilder1978} is based on the premise that there is a relationship between the number of times the price has gone up over a period $\Uparrow_{t}^{L}$ vs the number of times it has fallen $\Downarrow_{t}^{L}$ and assumes that the price is more likely to move upwards if $\Uparrow_{t}^{L}>\Downarrow_{t}^{L}$ and visa versa.

\end{itemize}

\subsubsection*{Calculation of p-values}
\begin{itemize}

\item For each in sample period, the proportion of occurrences of each of the three classes of movement (up, down or none) over the 100 instances of in sample data was determined. 

\item Predictions of movement were then generated randomly for each of the instances of the out of sample period where a prediction was deemed possible by SimpleMKL / individual kernel (as explained in section 3), each class having a probability of being assigned based on the in sample proportions.

\item This was repeated $10^5$ times for each out of sample section with the number of times the randomly generated predictions were correct along with the number of times SimpleMKL / individual kernel was correct for that period recorded each time. 

\item The proportion of the $10^5$ iterations that the number of correct predictions recorded for all the out of sample periods was greater than that reported by SimpleMKL / individual kernel was used to calculate the P-value.

\item In the work reported here, not one of the $10^5$ iterations of randomly generated predictions outperformed the SimpleMKL / individual kernel methods.

\end{itemize}

\end{document}